\documentclass[12pt]{article}
\usepackage[cp1251]{inputenc}
\usepackage{amstext}
\def\be{\begin{eqnarray}}
\def\ee{\end{eqnarray}}
\def\ba{\begin{array}}
\def\ea{\end{array}}

\begin{document}
\begin{center}
{\Large\bf Singularity-Free Interaction in  Dilaton-Maxwell\\ Electrodynamics} 
\end{center}

\begin{center}

O.V. Kechkin$^{[1],[2]}$ and P.A. Mosharev$^{[2]}$

\end{center}

\begin{center}
$[1]$ D.V. Skobeltsyn Institute of Nuclear Physics, M. V. Lomonosov Moscow State University, 119234, Moscow, Russia,\\
$[2]$ Faculty of Physics, M. V. Lomonosov Moscow State University, 119991, Moscow, Russia,
E-mail: kechkin@srd.sinp.msu.ru,\quad moscharev.pavel@physics.msu.ru

\end{center}

\begin{center}
{\Large\bf Abstract}
\end{center}
An effective potential is created for the dynamics of a test particle, which preserves dilatation symmetry for nonlinear static dilaton-Maxwell background. It is found that the central interaction in this theory is singularity-free everywhere; it vanishes at short distances and demonstrates Coulomb behavior far from the source. It is shown that static and spherically symmetric source behaves like a soliton: it has the finite energy characteristics that are inversely proportional to the dilaton-Maxwell coupling constant.
\begin{center}
\end{center}

\begin{center}
\end{center}

\section*{Introduction}

Nonlinear generalizations of Maxwell's electrodynamics are predicted by multidimensional Kaluza-Klein theories, and various supergravity and superstring theoretical schemes \cite{KK}--\cite{OK-1-0}. They can also be obtained using the most general algebraic approach, based on arguments about the hidden symmetries of the considered theory \cite{NED-1}--\cite{NED-3}. Main problems of standard linear electrodynamics, which can probably be resolved on non-linear way are related with different singularities predicted by the theory. For example, it is well known fact that the Coulomb interaction between two point charges diverges at small distances, as well as the energy of a single point charge. In this work we continue the study of dilaton-Maxwell electrodynamics (DME) -- the nonlinear theory, which occurs in various realizations of the Grand Unification Theory. We show that the theory, which possesses dilatation symmetry, is free of singularities mentioned above: effective interaction, as well as the energy of point sources prove to be the finite ones in non-linear variant of DME.

In \cite{OK-P} we studied the system with the dilatation symmetry that is realized on the dilaton-Maxwell background only: this symmetry was not supported by the dynamics of test charging particles. In this paper, we `extend'\, dilatation symmetry approach to the total scheme under consideration. We use the duality between the static dilaton-Maxwell  background and stationary General Relativity, which was found in \cite{OK-P}, to work with essentially nonlinear DME system. Namely, we perform the so-called `charging Ehlers transformation'\ (see \cite{Kinnersley-1}--\cite{OK-2-2}) to create the most general spherically symmetric dilaton-Maxwell background, which describes the fundamental electrical/magnetic sources in DME. We present the explicit form of the potential describing the interaction that is invariant under the action of dilatation symmetry and examine its properties. This interaction, as expected, demonstrates the Coulomb behavior far from the source, and 
it wonderfully has not any singularity at short distances.

\section{Dilatation symmetry in 4D dynamics}

Following \cite{OK-P}, let us consider the dilaton-Maxwell electrodynamics with negative kinetic term for dilaton field: \be\label{DME2-1}L_4=-\frac{1}{4}e^{-2\alpha\phi}F_{\mu\nu}F^{\mu\nu}- 2\partial_{\mu}\phi\,\partial^{\mu}\phi.\ee
Here
$F_{\mu\nu}=\partial_{\mu}A_{\nu}-\partial_{\nu}A_{\mu}$ is the conventional Maxwell's tensor, whereas  
$\phi=\phi\left(x^{\lambda}\right)$ and $A_{\mu}=A_{\mu}\left(x^{\lambda}\right)$
are the dilaton and electromagnetic fields, respectively. Then, space-time is understood as four-dimensional ($\mu,\,\nu,\,\lambda=0,\dots,3$) and flat (Minkowsky metrics is taken in the form  $\eta_{\mu\nu}={\rm diag}\left(+1,\,-1,\,-1,\,-1\right)$). Dilaton-Maxwell coupling constant $\alpha$ is considered as an arbitrary real parameter, it can be taken as nonnegative without loss of generality. Moreover, here we are interested in nonlinear dynamics, which corresponds to a non-zero value of this coupling. Thus, $\alpha>0$ in this work.  

The Lagrangian (\ref{DME2-1}) has a dilatation symmetry
\be\label{DME2-2}\phi\rightarrow\phi+\Lambda,\qquad A_{\mu}\rightarrow e^{\alpha\Lambda}A_{\mu},\ee
where $\Lambda={\rm const}$. Let us consider the dynamics of test particles in this dilaton-Maxwell background. It is easy to see that the modified Loretz force, which allows the symmetry of (\ref{DME2-1}) can be entered using four-dimensional notation as   
\be\label{DME2-3}\frac{dp^{\mu}}{ds}=qe^{-\alpha\phi}F^{\mu\nu}u_{\nu},
\ee
where $p^{\mu}=mu^{\mu}$ is the 4-momentum of a particle, $q,\,m={\rm const}$ are the charge and mass,  $u^{\mu}=dx^{\mu}/ds$ is a 4-velocity calculated at the 4-trajectory
$x^{\mu}=x^{\mu}\left(s\right)$. Then, four-dimensional line element is calculated as  $ds^2=\eta_{\mu\nu}dx^{\mu}dx^{\nu}$. In \cite{OK-P}, the dynamics was studied with standard Loretz force (right-hand side of Eq. (\ref{DME2-3}) does not contain the dilaton term $e^{-\alpha\phi}$ in this case). In the present work, the dynamics (\ref{DME2-3}) in the background (\ref{DME2-1}) is defined by effective electromagnetic tensor $\tilde F^{\mu\nu}=e^{-\alpha\phi}F^{\mu\nu}$. It can be obtained from a standard dynamics by substitution of the effective electric and magnetic fields $\vec E_{eff}= e^{-\alpha\phi}\vec E$ and $\vec H_{eff}= e^{-\alpha\phi}\vec H$  instead of the original quantities $\vec E$ and $\vec H$ in the corresponding dynamic relations. This new dynamics is the most symmetric: it is invariant under the transformation (\ref{DME2-2}) provided that the trajectory $x^{\mu}=x^{\mu}\left(s\right)$ is understood as an invariant of the dilatation symmetry.       

\section{Effective potential for static DME}

Let us consider the dynamics of test particles (\ref{DME2-3}) on a stationary background (\ref{DME2-1}), where the fields $A_{\mu},\,\phi$ do not depend on the time variable $x^0=t$ (these fields are functions of spatial variables $x^k$ ($k=1,\,2,\,3$)). In this case, one can use the magnetic pseudoscalar potential $\tilde A^0$ instead of the vector potential $\vec A=\left\{A^k\right\}$; it is defined by the relation $\nabla \tilde A^0=-e^{-2\alpha\phi}\nabla\times\vec A$ (where $\nabla=\left\{\partial_k\right\}$). 
Electrostatic (e) / magnetostatic (m) sectors of DME are defined as
\be\label{DME2-4}\left\{\ba{c}e\\m\ea\right\}:\qquad A^0= \left\{\ba{c}A\\0\ea\right\},\quad \tilde A^0= \left\{\ba{c}0\\A\ea\right\};\ee
dynamics (\ref{DME2-1}) for background fields corresponds to the following effective three-dimensional Lagrangian: 
\be\label{DME2-5}L_3=\frac{2}{\alpha^2}\,f^{-2}\left[\left(\nabla f\right)^2+\left(\nabla\chi\right)^2\right],\ee
where
\be\label{DME2-6}f=e^{\pm \alpha\phi},\qquad \chi=\frac{\alpha}{2}\,A\ee
in the theory with $\alpha\neq 0$ \cite{OK-P}. One can prove that Eq. (\ref{DME2-3}) in the static DME takes the form
\be\label{DME2-7}\frac{dP^0}{dt}=q\left\{\ba{c}\vec v\,\vec G\\0\ea\right\},\qquad \frac{d\vec P}{dt}=q\left\{\ba{c}\vec G\\\vec v\times\vec G\ea\right\},\ee
where 
$P^0=m/\sqrt{1-v^2}$ is the total (rest plus kinetik) relativistic energy, $\vec p=m\vec v/\sqrt{1-v^2}$ is the 3-momentum of the relativistic particle, and 
$\vec v=\left\{v^k\right\}$  (where $v^k=dx^k/dt$) is the standard 3-velocity. Then, 
\be\label{DME2-8}\vec G=-\frac{2}{\alpha}\,f^{-1}\nabla\chi=\left\{\ba{c}\vec E_{eff}\\\vec H_{eff}\ea\right\},\ee
i.e., the vector field $\vec G$ means effective electric /magnetic field of a static sector of DME under consideration.

From the Lagrangian (\ref{DME2-5}) it follows that the special ansatz $\chi_0=0$  $f_0=e^{\alpha\Upsilon}$ satisfies the corrersponding Euler-Lagrange system if $\Delta\Upsilon=0$, i.e. if `starting'\, dilaton field $\phi_0=\pm\Upsilon$ is harmonic. 
In \cite{OK-P} was presented a normalized Ehlers transformation in this theory, its application to this ansatz generates a background 
$f,\,\chi$ with
\be\label{DME2-9}f=\frac{f_0}{\cos^2\frac{\lambda}{2}+f_0^2\sin^2\frac{\lambda}{2}},\qquad \chi=-\frac{\sin\lambda}{2}\cdot
\frac{1-f_0^2}{\cos^2\frac{\lambda}{2}+f_0^2\sin^2\frac{\lambda}{2}},\ee
where $\lambda$ is an arbitrary real parameter. Substitution of Eq. (\ref{DME2-9}) to Eq. (\ref{DME2-8}) leads to the following result  for the electric/magnetic field $\vec G$: 
\be\label{DME2-10}\vec G=-\frac{2\sin\lambda}{\alpha}\cdot \frac{\nabla f_0}{\cos^2\frac{\lambda}{2}+f_0^2\sin^2\frac{\lambda}{2}}.\ee
The statement is that $\vec G$ is a potential field, $\vec G=-\nabla V$, where $V$ is the corresponding potential:
\be\label{DME2-11-1} V=\frac{4}{\alpha}\arctan\left[\frac{\tanh\left(\frac{\alpha\Upsilon}{2}\right)\sin\lambda}{1-\tanh\left(\frac{\alpha\Upsilon}{2}\right)\cos\lambda}\right]\ee
(here the integration constant is taken to have $V=0$ for $\Upsilon=0$). Note that the potential (\ref{DME2-11-1}) is finite. Thus, the dilaton field provides the regularization of the singularities at least for electro/magneto static backgrounds.
Then, the generated dilaton field is given by the relation:
\be\label{DME2-11-2}\phi=\mp\frac{1}{\alpha}\log\left[\cosh\left(\alpha\Upsilon\right) -\sinh\left(\alpha\Upsilon\right)\cos\lambda\right].\ee
It is easy to see that 
\be\label{DME2-11-3}V\rightarrow 2\Upsilon\sin\lambda+o\left(\Upsilon\right),\quad 
\phi\rightarrow \pm\Upsilon\cos\lambda+o\left(\Upsilon\right)\ee
if $\Upsilon\rightarrow 0$. For example, for multipole expansion  
\be\label{DME2-11-4}\Upsilon=\frac{Q_0}{r}+\dots\ee at $r\rightarrow \infty$ with nontrivial monopole charge $Q_0={\rm const}$ one obtains from Eq. (\ref{DME2-11-3}) that
\be\label{DME2-11-5}V=\frac{Q_{e/m}}{r}+\dots ,\qquad \phi=\frac{Q_d}{r}+\dots, \ee
where electric/magnetic charges $Q_{e/m}$ and dilaton charge $Q_d$ are given by the relations 
\be\label{DME2-11-6} Q_{e/m}=2Q_0\sin\lambda,\qquad Q_d=\pm Q_0\cos\lambda.\ee
It is interesting to note that the charge combination $\sqrt{Q_{e/m}^2+4Q_d^2}$ is invariant under the action of the charging symmetry transformation (\ref{DME2-9}), see \cite{OK-P} for details.
Then, in another intersting limit related to $\Upsilon\rightarrow+\infty \cdot \sigma$, where $\sigma=\pm 1$; the result reads as follows:
\be\label{DME2-11-44}V\rightarrow \frac{4\sigma}{\alpha}\arctan
\left(\frac{\sin\lambda}{1-\sigma\cos\lambda}\right)\ee
(and $\phi$ diverges in this case). 

The total energy of test particle is defined as 
\be\label{DME2-12}E=\left\{\ba{c}P^0+qV\\ P^0\ea\right\};
\ee
it is a motion integral for the dynamics under consideration, as follows from Eq. (\ref{DME2-7}). Here $U=qV$ is the potential energy of a test particle in the background fields; of course, it exists in electrosetatic sector only.

\section{Effective spherically symmetric sourse} 

Taking $\Upsilon=Q_0/r$, one obtains the most general dilaton-Maxwell background in spherically symmetric case \cite{OK-P}. The explicit form of the corresponding effective potential reads: 
\be\label{DME2-13}V=\frac{4}{\alpha}\arctan\left[
\frac{Q_{e/m}\tanh\left(\frac{\alpha\sqrt{Q_{e/m}^2+4Q_d^2}}{4r}\right)}{\sqrt{Q_{e/m}^2+4Q_d^2}\mp 2Q_d\tanh\left(\frac{\alpha\sqrt{Q_{e/m}^2+4Q_d^2}}{4r}\right)}\right],\ee
as it follows from Eq. (\ref{DME2-11-1}). Note, that Eq. (\ref{DME2-13}) gives a general modification of the Coulomb potential at the accepted values of the dilaton and electric/magnetic charges. Then, using Eq.(\ref{DME2-11-2}), we get  \be\label{DME2-14}\phi=\mp\frac{1}{\alpha}\log\left[\cosh\left(\frac{\alpha\sqrt{Q_{e/m}^2+4Q_d^2}}{2r}\right)\mp \frac{2Q_d}{\sqrt{Q_{e/m}^2+4Q_d^2}}\sinh\left(\frac{\alpha\sqrt{Q_{e/m}^2+4Q_d^2}}{2r}\right)\right]\ee for the dilaton field.
It is not difficult to prove that the asymptotics of Eqs. (\ref{DME2-13}) and (\ref{DME2-14}) at $r\rightarrow\infty$ are given by Eq. (\ref{DME2-11-5}), whereas 
\be\label{DME2-15}V\rightarrow \frac{4}{\alpha}\arctan\left(\frac{Q_{e/m}}{\sqrt{Q_{e/m}^2+4Q_d^2}\mp2Q_d}\right)\equiv V\left(0\right)\ee
in the limit $r\rightarrow 0$, which completely agrees with the general result (\ref{DME2-11-44}).

In the spherically symmetric case, the electric/magnetic field is radial, $\vec G=G\vec e_r$, where $\vec e_r$ is the unit radial vector, 
and
\be\label{DME2-16}G=\frac{Q_{e/m}}{r^2}\,f\ee
as follows from Eq. (\ref{DME2-9}). Here $G=-dV/dr$, thus
${\rm sign\left(dV/dr\right)}=-{\rm sign}\left( Q_{e/m}\right)$,
therefore, the function $V=V\left(r\right)$ decreases (increases) monotonically  for $Q_{e/m}>0$ ($Q_{e/m}<0$). From this fact it follows that 
$V\left(0\right)$ is the height of the potential barrier 
(depth of the potential well) in this problem. Then, from the analysis of Eq. (\ref{DME2-16}) it follows that 
$G\left(0\right)=G\left(\infty\right)=0$, i.e. the interaction vanishes at long and short distances. It is also possible to show that the function $G\left(r\right)$ reaches its extremal value (max (min) for $Q_{e/m}>0$ ($Q_{e/m}<0$)) at a distance   
\be\label{DME2-16-2}
r=r_{\star}=\frac{\alpha\sqrt{Q_{e/m}^2+4Q_d^2}}{2\xi},\ee  
where $\xi$ is the only root of the equation  
\be\label{DME2-17}e^{2\xi}=\eta\frac{\xi+2}{\xi-2}\ee
and
\be\label{DME2-18}\eta=\frac{\sqrt{Q_{e/m}^2+4Q_d^2}\pm 2Q_d}{\sqrt{Q_{e/m}^2+4Q_d^2}\mp 2Q_d}.\ee

Calculation of standard electric/magnetic energy in a spherically symmetric dilaton-Maxwell background gives the following finite result:
\be\label{DME2-19}E_{e/m}=\frac{1}{2}\int d^3x\,\vec G^2=\frac{4\pi Q_{e/m}^2}{\alpha\left(\sqrt{Q_{e/m}^2+4Q_d^2}\mp 2Q_d\right)}.\ee
Note that $E_{e/m}\sim 1/\alpha$,  i.e. the total energy is inversely proportional to the coupling constant, which is a typical property of a soliton.  Then, a special case with
$Q_d=0$  is the closest to the traditional dilaton-free background; for electric monopole one gets $E_e=4\pi \left|Q_e\right|/\alpha$ here. From Eq. (\ref{DME2-15}) we get $V_e\left(0\right)=\pi{\rm sign}\left(Q_e\right)/\alpha$ for the same system, and work $A$ that one needs to create the background, transferring the electric charge $Q=Q_e$ to the center of the field distribution from infinity is $A=\pi \left|Q_e\right|/\alpha$. Thus,  $A=E_e/4$, so
`energy of creation'\, is only a quarter of the total energy of the object, which can be calculated by examining the dynamics of test particles. This interesting fact corresponds precisely to the dark matter phenomenology. 

\section*{Conclusion}      

Thus, the dilatation symmetry can be used for constructing of nonlinear generalization of the standard Maxwell electrodynamics. The resulting theory is characterized by the finite and singularity-free interaction at all distances. In particular, a modified Coulomb's force vanishes at short distances; this means that dilaton field modifies the electromagnetic interaction so that it obtains some effective non-Abelian properties. In addition, the sources of the electromagnetic field in this theory demonstrate a number of properties that are characteristic of solitons.

It is important to emphasize that all non-Abelian properties of the static dilaton-Maxwell electrodynamics are related with the corresponding structures of stationary General Relativity due to the duality established between these systems in \cite{OK-P}. However, this duality does not mean complete equivalence of these theories. For example, there is no physical analogy for a Schwarzschild's black hole  in both versions of DME (developed in \cite{OK-P} and in this paper), although these DMEs contain a solution that relates to the Schwarzschild solution in General Relativity (see sec. 3 of the present work and the corresponding section in \cite{OK-P} for DME backgrounds that are related to the Taub-NUT solution via DME/GR duality). 

\section*{Acknowledgments} We thank our colleagues for encouraging.


\begin{thebibliography}{100}

\bibitem{KK}
J.M. Overduin, P.S. Wesson, Phys. Rept. {\bf 283},  303 (1997).

\bibitem{SS}
D. Youm, Phys. Rept. {\bf 316}, 1 (1999).

\bibitem{OK-1-0}
O.V. Kechkin, Phys. Part. Nucl. {\bf 35}, 383 (2004).

\bibitem{NED-1}
V.I. Denisov, V.A. Sokolov, M.I. Vasili'ev, Phys. Rev. {\bf D90}, 023011 (2014).


\bibitem{NED-2}
V.I. Denisov, Phys. Rev. {\bf D61}, 036004 (2000).

\bibitem{NED-3}
L. Labun, J. Rafelski, Phys. Rev. {\bf D81}, 065026 (2010).

\bibitem{OK-P}
O.V. Kechkin, P.A. Mosharev, `Structures of General Relativity in Dilaton-Maxwell Electrodynamics',  arXiv:1602.04634.



\bibitem{Kinnersley-1}
W. Kinnersley, J. Math. Phys. {\bf 18}, 1529 (1977).    


\bibitem{Ehlers}
J. Ehlers, in
`Les Theories Relativistes de la Gravitation',\, CNRS, Paris, 275 (1959). 

Ehlers, J. (1959). In Les Th?ories relativistes de la gravitation (Actes du Colloque Int., Royaumont, 21–27 Juin 1959) (CNRS, Paris).

\bibitem{OK-0}
O.V. Kechkin, Gen. Rel. Grav. {\bf 31}, 1075 (1999).

\bibitem{OK-1-1-1}
D.V. Gal'tsov, O.V. Kechkin, Phys. Lett. {\bf B361}, 52 (1995).

\bibitem{OK-2-1}
O.V. Kechkin, Phys. Rev. {\bf D65}, 066006 (2002). 

\bibitem{OK-2-2}
A. Herrera-Aguilar, O. Kechkin, Phys. Rev. {\bf D59}, 124006 (1999).  

\end{thebibliography}
\end{document}